\documentclass[aps,prb,superscriptaddress, twocolumn, showpacs]{revtex4-1}
\usepackage[utf8]{inputenc}
\usepackage[T1]{fontenc}
\usepackage{multirow}
\newcommand{\un}[1]{\ensuremath{\, \mathrm{#1}}}
\newcommand{\e}[1]{\ensuremath{\times 10^{#1}}}
\newcommand{\etal}{\emph{et al. }}

\newcommand{\vs}{\emph{vs. }}
\newcommand{\sq}{\ensuremath{\tilde{S}_Q}}
\newcommand{\hla}[1]{#1}
\newcommand{\hlb}[1]{#1}
\usepackage{graphicx}
\usepackage{subfigure}
\usepackage{amsmath}

\begin{document}
\title{Thermal properties of charge noise sources}
\author{Martin V. Gustafsson}
\email[]{martin.gustafsson@chalmers.se}

\author{Arsalan Pourkabirian}
\thanks{M.V.G. and A.P. contributed equally to this work.}

\author{Göran Johansson}
\affiliation{Microtechnology and Nanoscience, Chalmers University of
Technology, SE-41296, Göteborg, Sweden}
\author{John Clarke}
\affiliation{Microtechnology and Nanoscience, Chalmers University of
Technology, SE-41296, Göteborg, Sweden}
\affiliation{Department of Physics, University of California, Berkeley, California 94720-7300, USA}
\author{Per Delsing}
\email[]{per.delsing@chalmers.se}
\affiliation{Microtechnology and Nanoscience, Chalmers University of
Technology, SE-41296, Göteborg, Sweden}

\date{\today}

\begin{abstract}\hlb{Measurements of the temperature and bias dependence of Single Electron Transistors (SETs) in a dilution refrigerator show that charge noise increases linearly with refrigerator temperature above a voltage-dependent threshold temperature, and that its low temperature saturation is due to SET self-heating. We show further that the two-level fluctuators responsible for charge noise are in strong thermal contact with the electrons in the SET, which can be at a much higher temperature than the substrate. We suggest that the noise is caused by electrons tunneling between the SET metal and nearby potential wells.}\end{abstract}

\pacs{73.50.Td, 85.35.Gv, 03.65.Yz, 65.80.-g}

\maketitle


\section{Introduction}
Low-frequency charge noise with a power spectral density $S_Q(f) \propto 1/f^\alpha$ is observed in all charge sensitive devices ($f$ is frequency and $\alpha \approx 1$). Apart from limiting the sensitivity of electrometers, such as the Single Electron Transistor (SET) \cite{Averin1991, Fulton1987}, charge noise is a source of decoherence in qubits \cite{Paladino2002, Bylander2011} and gives rise to errors in metrological quantum standards \cite{Kautz2000, Covington2000,  Bylander2005}. The noise is usually attributed to a superposition of Lorentzian spectra, each generated by a Two-Level Fluctuator (TLF) consisting of a charged particle moving stochastically between two locations\cite{Machlup1954, Dutta1981}. In spite of extensive studies, both experimental and theoretical \cite{Dutta1981, Zimmerli1992, Zorin1996, Zimmerman1997, Brown2006, Kafanov2008, Wolf1997, Song1995, Li2007, Henning1999, Krupenin2000, Kenyon2000, Astafiev2006, Starmark1999}, the sources and locations of the TLFs remain unknown.

\begin{figure}
\includegraphics{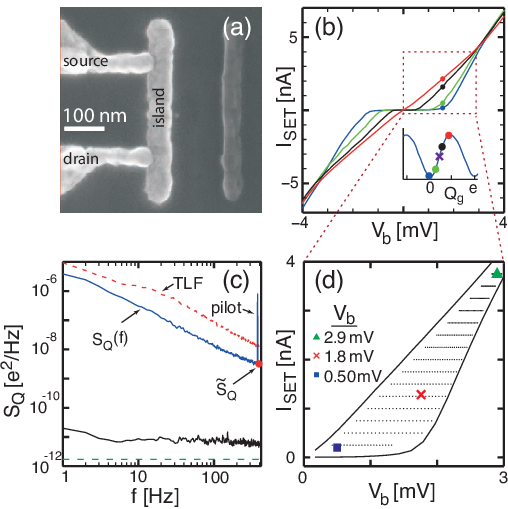}
\caption{\label{device_and_spectra_and_biaspoints}Single Electron Transistor. (a) Scanning electron microscope image showing the SET design. The area of each junction is approximately $60 \times 30 \un{nm^2}$ The island to the far right is a result of the two-angle evaporation, and plays no role in the SET operation. The gate electrode is located outside the picture, $600\un{nm}$ to the right of the island. (b) $I$-$V$ characteristics of device S1, taken at four different gate voltages. Inset: Typical measured $I_{SET}$ \vs $Q_g$, the charge induced from the gate. The colored points correspond to those in the $I$-$V$ plot. The cross denotes the operating point used in \emph{Exps. A} and \emph{B}. (c) Solid blue curve: Typical noise power spectrum $S_Q(f)$ acquired at $T_{0}=50 \un{mK}$. The slope of this spectrum is $\alpha = 1.24\pm 0.01$. The pilot signal, used for gain calibration, appears as a strong peak at $f_p = 377\un{Hz}$. The red dot marks $\sq$, the average of $S_Q$ between $383\un{Hz}$ and $401\un{Hz}$. The dashed green line is the shot noise of the SET for this particular value of $I_{SET}$. The solid black curve shows the amplifier noise, measured with open input, scaled to units of charge by the charge gain of the SET. The dashed red curve is an example of a spectrum (at $T_{0}=1.5 \un{K}$) with a substantial contribution from a single TLF at intermediate frequencies (see text). (d) Solid lines: Maximum and minimum SET current as a function of bias voltage. Blue square, red cross and green triangle: Bias points used for the temperature sweep of \emph{Exp.\,B}. \hlb{At fixed $V_b$, the gate voltage $V_g$ was varied to adjust $I_{SET}$ to the midpoint between the maximum and minimum SET current [1(b) inset]}.
Dots: Measurement points where the charge noise was measured (at base temperature) in \emph{Exp.\,C}. \hlb{The bias voltage was fixed, and $I_{SET}$ subsequently adjusted to the desired value by varying $V_g$.}}
\end{figure}

Because of its simplicity and unmatched charge sensitivity, the SET is an ideal tool to study the fundamentals of charge transport and noise\cite{Xue2009, Ubbelohde2012, Bylander2005}, and its structural similarity to superconducting qubits and metrological devices means that knowledge gained from the SET can be carried over to these and other devices. The SET consists of a small metallic island connected via tunnel junctions to source and drain electrodes [Fig.\,\ref{device_and_spectra_and_biaspoints}(a)]. Its current-voltage ($I$-$V$) characteristic [Fig.\,\ref{device_and_spectra_and_biaspoints}(b)] depends strongly on the charge induced on the island by externally applied electric fields or charges moving in its surroundings, and this response is periodic in the electron charge $e$ [Fig.\,\ref{device_and_spectra_and_biaspoints}(b), inset]. 

The electrons moving through the SET deposit energy on the island in proportion to the bias voltage $V_b$. To a good approximation, the power dissipated on the SET island is $P_{SET} = V_b I_{SET}/2$, and in a metallic SET this power is generally assumed to be dissipated rapidly into the electron gas. At low temperatures, the thermal coupling between the SET electrons and the phonons of the substrate is weak, and the temperature of the electron gas can be elevated significantly above that of the phonons\cite{Kautz1993}. The effect of this on the charge noise is discussed further below.

Several authors have found that the charge noise measured in SETs decreases as the temperature is lowered, saturating to a constant level at low temperature \cite{Song1995, Li2007, Henning1999, Krupenin2000, Kenyon2000, Astafiev2006}. This saturation has been attributed to self-heating of the SET but, since there is no obvious model for thermalization between the TLFs and the SET electron gas, the issue remains open.

The influence of SET bias parameters on charge noise has also been studied \cite{Wolf1997, Henning1999, Krupenin2000}, and while it is clear that the charge noise increases with SET bias, no quantitative conclusions have been drawn. This is largely due to the small range of useful bias voltages in SETs with modest charging energy, and the tendency of noise data to suffer from scattering and drift.

Here, we present three experiments on charge noise in SETs. Using SETs with relatively high charging energy and recording an extensive amount of data, we can clearly resolve the dependence of the noise on both refrigerator temperature, $T_0$, and SET bias parameters. \hlb{Our results demonstrate that the TLFs predominantly thermalize with the electron gas of the SET, rather than by a roundabout route through the substrate phonons. At high $T_0$, the SET electrons (and hence the TLFs) are thermalized with the refrigerator temperature, so that the charge noise scales proportionally with $T_0$. For low $T_0$, self heating brings the SET electrons to a temperature $T_e > T_0$, which causes the charge noise to saturate as $T_0$ is decreased}. This is expected if the TLFs are occupied by electrons tunneling from the SET metal, but not for conventional double-well TLFs located entirely outside the SET. Thus, our data point to a scenario where the noise is dominated by single-well TLFs close to the SET junctions.

\section{TLF models}
\begin{figure}
\includegraphics{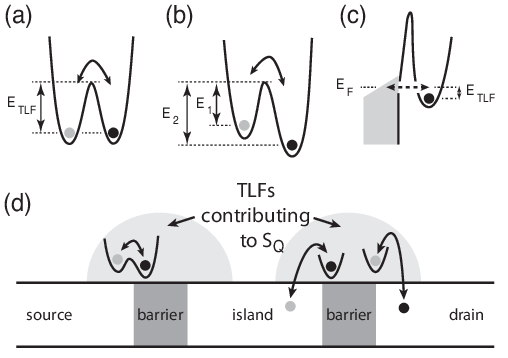}
\caption{\label{TLF_sketch}Different models for Two-Level Fluctuators. (a) The TLFs that produce charge noise are usually modeled as charged particles, each moving stochastically between two potential wells in the dielectrics surrounding the SET. With symmetrical wells, and a single activation energy $E_{TLF}$ which the particle must overcome by thermal excitation, the model predicts\cite{Dutta1981} $S_Q \propto T/f$. (b) With two defining energy scales ($E_1$ and $E_2$), the model instead predicts\cite{Kenyon2000} $S_Q \propto T^2/f$. (c) The TLF model suggested by our experimental data consists of a single potential well which may be occupied (unoccupied) by an electron tunneling from (to) the SET electrodes. There should be approximately as many wells coupled to the leads as to the island, and both types of well contribute equally to the charge noise, assuming uniform temperature. $E_{TLF}$ is the depth of each TLF with respect to the Fermi energy $E_F$ of the metal to which it is tunnel-coupled. \hlb{This model predicts a $S_Q \propto T/f^\alpha$ dependence (see text)}. (d) Only TLFs located inside the junctions or within a few barrier thicknesses of them can change the charge induced on the SET island sufficiently to contribute to the observed noise.}
\end{figure}

Regardless of their microscopic origins, TLFs can affect the SET only if they are located inside or in close vicinity to the tunnel junctions [Fig.\,\ref{TLF_sketch}(c)]. Based on the dependence of the charge noise on external electric fields, for a sample similar to ours, Zimmerman \etal were able to prove that the TLF ensemble must be at least partly located outside the tunnel barriers\cite{Zimmerman1997}.

The most common model for a charge TLF is that of a charged particle moving stochastically back and forth between two potential wells located in the dielectric material surrounding the device [Fig.\,\ref{TLF_sketch}(a)]. For each of the two wells, a different fraction of the TLF charge couples to the device, in the case of an SET as charge induced on its island. \hlb{In the simplest model of a symmetric, two-well potential\cite{Dutta1981}, switching between states occurs with a characteristic time $\tau$ that is identical for the two directions. The resulting power spectrum is a Lorentzian of the form $L(\omega) \propto \tau/(1+\omega^2 \tau^2)$; here $\omega = 2 \pi f$. If, further, the process is thermally activated, $\tau = \tau_0 \exp (E/k_B T)$, where $1/\tau_0$ is a characteristic attempt frequency and $E$ is the barrier height. Thus, for an ensemble of TLFs with a distribution of energies $D(E)$, the noise power spectrum becomes $S(\omega, T) \propto \int L(E, \omega) D(E) dE$ . The function $L(E, \omega)$ is strongly peaked with a width of the order of $k_B T$. Finally, provided $D(E)$ varies slowly on the scale of $k_B T$, Dutta and Horn\cite{Dutta1981} show that $S(\omega, T)\propto (k_B T/\omega)D(\tilde{E})$, where $\tilde{E}$ is the energy at which $L(E, \omega)$ peaks. This result demonstrates that for an ensemble of symmetric charge TLFs, $S_Q\propto T/f$.

Kenyon \etal \cite{Kenyon2000} extended this picture to an asymmetric double-well potential [Fig. 2(b)] in which each TLF has two different activation energies, one for each direction of switching. Under the assumptions that the two switching processes are independent and the two energies are uniformly distributed over the ensemble of TLFs, each process contributes a factor scaling as $T$ to the power spectrum, leading to $S_Q \propto T^2/f$.}

It has also been proposed that each TLF consists of a single potential well, with electrons tunneling back and forth between the well and the conductors of the device\cite{Brown2006, Kafanov2008, Simkins2009} [Fig.\,\ref{TLF_sketch}(c)]. A TLF of this type can be tunnel-coupled either to the SET island or to one of the leads, with the two cases influencing the SET equally. \hlb{For an electron entering a well from the island (lead), the major part of its charge is induced back on the island (lead), and only a small fraction of its charge is induced on the lead (island)}. In both cases, the result is a small change in the offset charge of the island.

Potential wells of this type must reside within tunneling distance of the SET, and could consist of metal grains formed during device fabrication\cite{Brown2006, Kafanov2008} or surface states in the interface between the conductors and their surrounding oxides \cite{Simkins2009}. States of the latter type, known as Metal-Induced Gap States (MIGS)\cite{Louie1976}, exist at all disordered interfaces between conductors and insulators, with an areal density of around $0.5/\un{nm^2}$, and, when localized\cite{Choi2009}, have been suggested to harbor the electrons that produce magnetic flux noise in SQUIDs\cite{Clarke2004} and flux-sensitive qubits\cite{Paladino2002, Bylander2011}.

Assuming that the well and the conductor are in equilibrium at a temperature $T$, the rates for tunneling into ($\Gamma_{in}$) and out of ($\Gamma_{out}$) the well obey detailed balance and are related by
\begin{equation*}
	\Gamma_{out} = \Gamma_{in}\frac{P}{1-P} = \Gamma_{in}e^{-E_{TLF}/k_B T} ,
\end{equation*}
where $P$ is the probability of the well being occupied and $E_{TLF}$ the depth of the well with respect to the Fermi energy $E_F$ of the metallic reservoir. Each TLF of this type produces noise with a Lorentzian spectrum
\begin{equation*}
	L \propto \frac{\omega_0}{\omega_0^2+\omega^2} P(1-P) = \frac{\omega_0}{\omega_0^2+\omega^2} \left[\cosh \left( \frac{E_{TLF}}{2 k_B T}\right)\right]^{-2} ,
\end{equation*}
where the characteristic frequency of the Lorentzian is given by $\omega_0 = \Gamma_{in}+\Gamma_{out}$. Assuming, as for the other TLF models mentioned, that the activation energies $E_{TLF}$ are uniformly distributed over the ensemble of TLFs, the noise generated by the ensemble depends on temperature and frequency \hlb{as $S_Q \propto T/f^\alpha$. If the tunnel barriers between potential wells and metal have uniformly distributed thicknesses and heights, the model predicts $\alpha=1$. However, distributions that deviate from uniform produce $\alpha \ne 1$ without affecting the temperature dependence of the noise}. As we shall see, this is consistent with our experimental results.

\section{Experiments}
We fabricated several SETs using two-angle evaporation\cite{Fulton1987} on a single-crystalline silicon substrate covered with 400 nm of thermal oxide [Fig.\,1(a)], and included two of them, S1 and S2, in this study. The chip was cooled in a dilution refrigerator with a base temperature of $20\un{mK}$ and fitted with extensive low-temperature filtering of the measurement lines. A magnetic field of $0.6\un{T}$ quenched superconductivity in the aluminum, and the SET was voltage biased symmetrically with two nominally identical, home-built transimpedance amplifiers. \hlb{By varying the SET gate voltage $V_g$ for fixed bias voltage $V_b$, we adjust the current $I_{SET}$ to a working point appropriate to the measurement. In some measurements, $I_{SET}$ is chosen to give the highest sensitivity to charge noise, whereas in others $I_{SET}$ is chosen to produce a certain power dissipation in the SET (see below)}. By fitting $I_{SET}(V_b)$ and $I_{SET}(V_g)$ curves to numerical simulations we found that the SETs have charging energies of $E_{C,1}/k_B=10.6\un{K}$ and $E_{C,2}/k_B=6.0\un{K}$, and total resistances (sum of the two junction resistances at high bias) of $R_{\Sigma,1}=354\un{k\Omega}$ and $R_{\Sigma,2}=147\un{k\Omega}$. Typical $I$-$V$ characteristics for S1 are shown in Fig.\,1(b).

The experiment was divided into three parts, henceforth referred to as \emph{Exps. A, B} and \emph{C}. In all three cases, we measured charge noise in a frequency range from 1Hz to 401Hz \hla{using a Stanford Research Systems spectrum analyzer SR785}. A ``pilot'' tone at frequency $f_p = 377\un{Hz}$ with an accurately known charge amplitude was used to calibrate the noise level of the spectra. We extract a single value $\sq$ to represent the noise level of each acquired spectrum by averaging $S_Q$ over frequencies between $383\un{Hz}$ and $401\un{Hz}$ (above $f_p$).
Studying the noise at this relatively high frequency minimizes error due to the limited measurement time of each spectrum, and produces a low spread between neighboring temperature and bias points. \hla{In some spectra, a single TLF with abnormally strong coupling to the SET and low characteristic frequency is seen as a Lorentzian superimposed on the $1/f^\alpha$ background [dashed red line in Fig. 1(c)]. By extracting $\sq$ at relatively high frequency, we minimize the impact of this individual TLF}. We correct for two spurious noise sources. The SET shot noise\cite{Korotkov1994noise, Kafanov2009}, $S_S = e I$, is shown as a green dashed line in Fig.\,1(c). A typical noise power spectrum of the amplifier, scaled by the charge power gain of the SET and measured with open input, is plotted as a solid black line. The plotted spectrum is scaled by the charge gain of the SET. The value of $\tilde{S}_Q$ exceeds both noise levels by at least an order of magnitude, but we nonetheless subtract these two contributions from each spectrum in the post-processing to improve the accuracy of the data.

\hla{In separate time-domain measurements on nominally identical samples we find, as expected\cite{Weissman1988}, that the noise is nongaussian}.

\subsection{Temperature dependence of the noise}

In the first part of the study, \emph{Exp.\,A}, we measured $\sq$ for devices S1 and S2 (in separate runs) while increasing $T_0$ from $50 \un{mK}$ to around $4 \un{K}$ over a period of 18 to 19 hours. \hlb{We biased each SET at the voltage $V_b$ where its charge modulation is at a maximum}. Immediately before each noise spectrum measurement, we acquired a gate modulation trace [Fig.\,\ref{device_and_spectra_and_biaspoints}(b), inset], extracted the minimum and maximum current, and set the current at the midpoint between these values by adjusting the gate voltage. For each SET, we acquired data at $\sim 200$ temperature points, \hla{with each spectrum averaged 250 times}.

\begin{figure}
\includegraphics{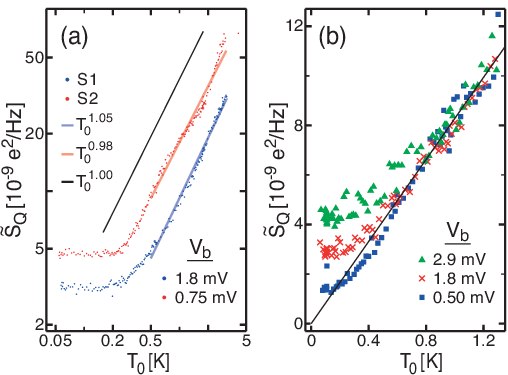}
\caption{\label{noise_vs_T}Charge noise $\sq$ as a function of refrigerator temperature $T_{0}$. (a) \sq\, for two SETs at fixed bias voltage (\emph{Exp.\,A}): S1 (lower, blue dots) and S2 (upper, red dots). The solid black line is a guide to the eye, illustrating linear temperature dependence. \hlb{The blue and red lines are linear fits in the logarithmic plot for temperatures above $0.5\un{K}$, for S1 and S2, respectively}. (b) $\tilde{S}_Q$ for S1, with bias voltage $V_b$ alternating between three different values (\emph{Exp.\,B}). The solid line is a linear fit through zero to the noise for $T_0 > 1\un{K}$; see text. Symbols correspond to those in Figs.\,\ref{device_and_spectra_and_biaspoints}(d) and \ref{noise_vs_bias}(c).}
\end{figure}

The noise data, displayed in Fig.\,\ref{noise_vs_T}(a), clearly show that $\sq$ increases linearly with temperature\hlb{: The slopes in the logarithmic plot are $1.05\pm0.02$ and $0.98\pm0.04$ for S1 and S2, respectively}. Other samples, not presented here, showed similar linear scaling with $T_0$. This scaling is in contrast with the $S_Q \propto T_0^2$ dependence presented by Kenyon \etal \cite{Kenyon2000} and Astafiev \etal \cite{Astafiev2006}, although the latter group has observed $S_Q \propto T$ in other devices \cite{Astafiev_Sq_propto_T}.

At temperatures below about $0.25\un{K}$ we observe a saturation of the noise, as reported by previous authors \cite{Wolf1997, Song1995, Li2007, Henning1999, Krupenin2000, Kenyon2000, Astafiev2006}. The noise level becomes completely independent of $T_0$ in this regime to within our measurement precision; see discussion below.

\subsection{Bias and temperature dependence of the noise}
In \emph{Exp.\,B}, we repeated the temperature sweep of \emph{Exp.\,A} for device S1 from $T_{0} = 50\un{mK}$ to $T_{0} = 1.5\un{K}$ over a period of 8 hours, while alternating the bias voltage $V_b$ of the SET between the three values shown with colored symbols in Fig.\,\ref{device_and_spectra_and_biaspoints}(d). \hla{For each data point, the spectrum was averaged 100 times}.\,The noise data, plotted in Fig.\,\ref{noise_vs_T}(b), clearly show that the saturation levels at low $T_0$ depend on the bias voltage, which is consistent with a picture of TLFs activated by hot electrons in the SET.

As an alternative to SET self-heating, it has been proposed that electrons tunneling through the junctions supply energy directly to the TLFs, either by scattering inelastically with TLFs located inside the junction barriers\cite{Kenyon2000}, or by coupling to the electric field generated by the SET shot noise\cite{Wolf1997}. In such processes, the tunneling electrons would be able to activate TLFs with energies up to $e V_b$, and a much larger number of TLFs would be activated for the highest value of $V_b$ than for the lowest one. This difference should persist as $T_0$ increases, until all TLFs can be thermally activated. On the contrary, we see that the bias dependence of TLF activation vanishes at $T_0 \approx 0.6 \un{K} \ll e V_b/k_B$ [Fig.\,\ref{noise_vs_T}(b)]. We conclude from this result that such direct activation mechanisms cannot explain the bias dependence of \sq.

In the regime $T_0 \ge 1 \un{K}$, we see a clear linear dependence $\sq = \beta T_0$ with $\beta = (8.28 \pm 0.02)\e{-9}\un{e^2 Hz^{-1} K^{-1}}$. Using this relation, as we see in the following section, we can calculate the saturation temperature $T_{TLF}$ of the TLF ensemble for each of the three bias points.

\subsection{Bias dependence of the noise at base temperature}
To investigate in greater detail the connection between charge noise and bias voltage and current found in \emph{Exp.\,B}, we performed \emph{Exp.\,C} for S1, in which we measured the low-temperature saturation level of the charge noise at 315 bias points, each with a different value of $V_b$ and $I_{SET}$ [Fig.\,\ref{device_and_spectra_and_biaspoints}(d)]. \hla{Each spectrum was averaged 100 times.} The total measurement time was 12.6 hours, and the bias points were applied in random order to avoid any influence of measurement drift on the data.

\begin{figure}
\includegraphics{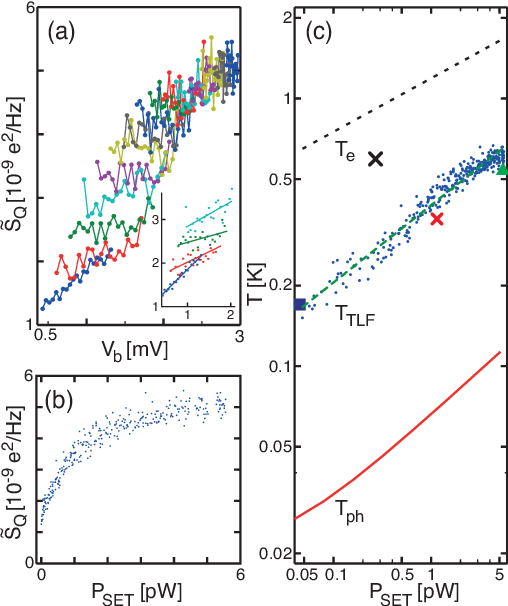}
\caption{\label{noise_vs_bias}Noise and temperature \vs SET bias for S1, at refrigerator base temperature (\emph{Exp.\,C}). (a) Charge noise $\sq$ \vs $V_b$. Points grouped by color and connected with lines were acquired with the same nominal value of $I_{SET}$. Inset: Data for the lowest four values of $I_{SET}$ with linear fits. For low values of $V_b$ (traces with the lowest $\sq$), there is a clear increase in noise level with bias voltage. (b) The same $\sq$ data as in (a) plotted versus power dissipated in the SET, $P_{SET}= V_b I_{SET}/2$. (c) Extracted TLF temperature $T_{TLF}$ (points) versus $P_{SET}$. \hla{The central line (dashed green) is fitted to the data the logarithmic plot, yielding $T_{TLF} \propto P_{SET}^{0.29\pm0.01}$}. The square, cross and triangle are the noise saturation levels for the three different bias points of \emph{Exp.\,B}. The upper dashed line (black) is the calculated electron temperature of the SET island, and the black cross is the electron temperature extracted from the modulation curve of the SET [see text and Fig. 5(b)]. The lower solid line (red) is the calculated phonon temperature beneath the SET (see text).}
\end{figure}

Figure\,\ref{noise_vs_bias}(a) shows the charge noise level as a function of SET bias voltage $V_b$, with lines connecting points with the same bias current $I_{SET}$. It is clear from this plot that $\sq$ increases with both $V_b$ and $I_{SET}$.

Plotting the noise data of Fig.\,\ref{noise_vs_bias}(a) versus $P_{SET}$ instead of $V_b$, we see a smooth, monotonic increase in the noise with $P_{SET}$, but with a dependence much weaker than linear [Fig.\,\ref{noise_vs_bias}(b)]. This indicates that the TLFs are heated by the power dissipated in the SET. The weak power law is characteristic of electron-phonon thermalization, which is generally assumed to explain the power dependence of the electron temperature in the SET island\cite{Kautz1993, Korotkov1994heating, Verbrugh1995, Meschke2004}. 

In Fig.\,\ref{noise_vs_bias}(c), we have used the proportionality constant $\beta$ determined in \emph{Exp.\,B} to calculate the equivalent temperature of the TLFs, $T_{TLF}$, using the same noise data as in Figs.\,\ref{noise_vs_bias}(a) and (b). \hla{The data are fitted with a line in the logarithmic plot [Fig. 4(c), dashed green] to yield $T_{TLF} \propto P_{SET}^{0.29\pm0.01}$}. The three saturation temperatures extracted from \emph{Exp.\,B} are plotted on the same scale. \hlb{Since \emph{Exps.\,B} and \emph{C} were carried out more than one week apart, it is not unreasonable to expect the noise sources to have reconfigured somewhat, as commonly seen in experiments on low-frequency noise}. Nonetheless, the two data sets agree rather well. 

\section{Thermal modeling}
The current flowing through a SET dissipates power in the electron gas of both the island and the leads. The electrons on the island generally relax rapidly to a Fermi distribution, and these hot electrons may subsequently thermalize via preferential tunneling from the island and by emission of energy as phonons. 

Since this self-heating affects the SET current in a theoretically predictable way, we can extract a value for the electron temperature of the island from the $I$-$V$ characteristics of the SET.
\begin{figure}
\includegraphics{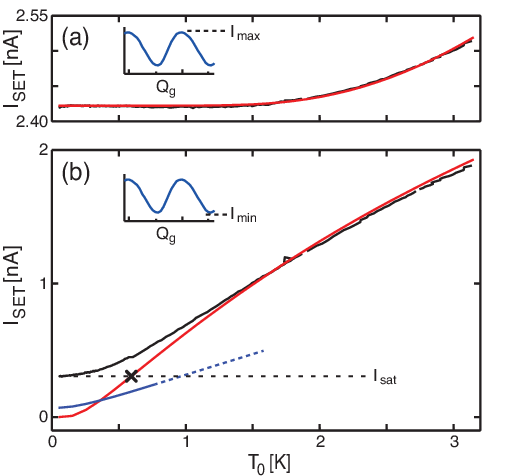}
\caption{\label{set_fitting}Extraction of the electron temperature from the $I$-$V$ characteristics of S1. (a) Maximum SET current versus refrigerator temperature, $T_0$, for fixed $V_b$. The black curve represents experimental data and the red curve is a fit to the standard (Orthodox) theory of the SET, for the regime $T_0>2\un{K}$ where this model is accurate. The bias voltage $V_b$ and the charging energy $E_C$ were used as fitting parameters. (b) Minimum SET current versus refrigerator temperature for the values of $V_b$ and $E_C$ found in (a). The solid black curve shows experimental data, and the red curve is the theoretical (Orthodox) calculation, with no fitted parameters. The experimental curve saturates at a value $I_{sat}$ which is higher than the theoretical one, as a result of SET self-heating. The blue curve is a theoretical calculation with co-tunneling included\cite{KonigThesis}, from which we can rule out that co-tunneling is the reason for the high value of $I_{sat}$. The cross-over between the experimental and Orthodox curves is marked with a black cross, and provides an experimental value for $T_e$. This data point is also plotted in Fig.\,\ref{noise_vs_bias}(c).}
\end{figure}
Along with each noise spectrum acquired in \emph{Exp.\,A}, we measured the gate modulation curve of SET S1 [Fig.\,\ref{device_and_spectra_and_biaspoints}(b), inset]. Both the maximum and minimum of each such curve ($I_{max}$ and $I_{min}$) depend on the electron temperature $T_e$, as shown in Fig.\,\ref{set_fitting}.

By fitting the standard (Orthodox) model for the SET model\cite{Averin1991} to $I_{max}$ versus $T_0$, we can accurately extract the charging energy $E_C$ of the SET, as well as the exact value of the bias voltage $V_b$. The extracted value for $E_C$ agrees with that obtained from $I$-$V$ characteristics measured at base temperature. Using these parameters, we can compare the temperature dependence of $I_{min}$ with the Orthodox model. We find that the data and theory agree well for high $T_0$, but that $I_{min}$ saturates  at low $T_0$, to a value $I_{sat}$ which is substantially higher than theory predicts. We also calculate the total current using the analytical method of König\cite{KonigThesis}. This treatment includes co-tunneling, but only considers two charge states on the SET island. Thus, this method is valid only for low values of $T_e$ and $V_b$, and in this regime we find that the co-tunneling current is much too low (by a factor of $\sim 8$) to account for $I_{sat}$. The cross-over temperature between the experimental curve and the Orthodox curve provides the experimental value $T_e = 0.59\un{K}$ at $P_{SET} = V_b I_{sat}/2$, as shown in Fig.\,\ref{set_fitting}(b). This data point is also shown as a cross in Fig.\,\ref{noise_vs_bias}(c).

Widely used models predict the electron-phonon thermalization power to follow $P_{e-ph}=\Sigma_n \Omega (T_{e}^n-T_{ph}^n)$ \hlb{with $n$ ranging approximately between 4 and 6, depending on geometry, temperature, and material properties\cite{Wellstood1994, Meschke2004, Schmidt2004a, Savin2006, Underwood2011}}. In this equation,  $\Sigma_n$ is a material-dependent electron-phonon coupling coefficient, $\Omega$ is the volume of the electron gas, and $T_{e}$ and $T_{ph}$ are the temperatures of the electrons and the phonons, respectively. Using\cite{Meschke2004} $n = 5$ and $\Sigma_5 = 0.4 \times 10^9 \un{W\,K^{-5}\,m^{-3}}$ for Al, we obtain an estimate of $T_{e}$ over the whole range of applied SET power [Fig.\,\ref{noise_vs_bias}(c)]. \hlb{The experimental data point [black cross in Fig. 4(c)] is within a factor 1.6 of the theoretical model [black dashed line in Fig. 4(c)], and we attribute the discrepancy to the relative simplicity of the electron-phonon thermalization model and the uncertainties in its input parameters}. Qualitatively, we see that $T_{TLF}$ and the theoretical $T_{e}$ have similar dependencies on $P_{SET}$.

We observe no increase in $\tilde{S}_Q$ at fixed SET bias for $50\un{mK} < T_0 \lesssim 250\un{mK}$ [Fig.\,\ref{noise_vs_T}(a)]. At these low temperatures, we expect $T_e$ on the SET island to be dominated by self-heating, while the electron gases of the leads are commonly believed to have temperatures close to\cite{Kautz1993} $T_0$. Geometrically, about as many of the TLFs contributing to $S_Q$ are located adjacent to the SET leads as to the island [Fig.\,\ref{TLF_sketch}(d)]. On this assumption, one-half of the TLFs (those that predominantly thermalize with the leads) should contribute charge noise proportional to $T_0$. In the regime of low $T_0$, these TLFs would produce a slope in $\tilde{S}_Q$ \emph{vs} $T_0$ of $\beta/2$. It is clear from Fig.\,\ref{noise_vs_T}(a) that this is not the case in our experiment. The electrons close to the junctions appear to follow the electron temperature on the SET island. We attribute this to local heating of the electrons in the leads closest to the junctions. This is plausible since the power dissipation, $P_{SET}/2$, in each of the leads takes place close to the junctions in comparison with the distance over which the electrons thermalize.

The electron-phonon coupling is usually assumed to be the dominant thermalization bottleneck for the island electrons, so that $T_e > T_{ph} \approx T_0$, and we used this approximation to calculate the electron temperature above. Nonetheless, some experiments have shown that the thermal power flowing from a SET with $T_e \gg T_0$ can produce a measurable increase in temperature of devices deposited nearby on the same substrate \cite{Krupenin1999, Savin2006}. Since we have no means to measure $T_{ph}$ in the region around the SET, we calculate an estimate from a finite-element model. The model is axially symmetric, with the SET defined as a disc at the surface with the same area as the actual SET, and with the same layer structure as the actual substrate. We assume a refrigerator temperature $T_0 = 20\un{mK}$ and use established literature values for the temperature-dependent thermal conductivities of $\Lambda_\mathrm{Si} = 5.0 T^3/{\un{K}^3} \un{W K^{-1} m^{-1}}$ for\cite{Kumar1985} Si and $\Lambda_\mathrm{SiO_2} = 0.03 T^2/{\un{K}^2} \un{W K^{-1} m^{-1}}$ for\cite{Pohl2002} $\mathrm{SiO_2}$, respectively. We assume that all the power $P_{SET}$ dissipated in the SET island is emitted as phonons from the Al/$\mathrm{SiO_2}$, and treat the materials as bulk media. We find that the calculated $T_{ph}$ at the hottest point in the model is much too low to account for the elevated temperature of the TLF ensemble [Fig.\,\ref{noise_vs_bias}(c)]. The only mechanism for the TLFs to assume a higher temperature than the local phonons is via direct contact with the SET electrons.

\section{Conclusions}

Analysis of our noise data yields new information on the processes responsible for charge noise in mesoscopic devices. We summarize the results of our investigation in four conclusions:

\hlb{(\emph{i}) We see clearly that the charge noise increases linearly with refrigerator temperature for high temperatures, and saturates for low temperatures to a value that depends on the SET bias.

(\emph{ii}) In the regime of low refrigerator temperature, the dependence of the charge noise on SET bias voltage and current is compelling evidence that the TLF ensemble dominating the noise is activated by SET self-heating.

(\emph{iii}) By our estimates, the temperature of the TLF ensemble is approximately five times higher than the local surface temperature of the substrate, and two to three times lower than the electron temperature of the SET, at refrigerator base temperature [Fig.\,\ref{noise_vs_bias}(c)]. This indicates that the TLFs are in  stronger thermal contact with the SET electrons than with the phonons in the substrate.}

(\emph{iv}) As Kenyon \etal have pointed out\cite{Kenyon2000}, it is difficult to see how double-well TLFs outside the SET would thermalize with the SET electron gas. On the contrary, a process by which the noise is generated by electrons tunneling between the SET island and local defects (such as localized MIGSs or metallic grains) would account naturally for this thermal coupling, \hlb{and would also produce the $S_Q \propto T/f^\alpha$ dependence we observe [Fig.\,\ref{TLF_sketch}]}. Since the electron shares its time between the SET metal and the external well, it is reasonable that the fluctuator has an equivalent temperature lower than that of the SET electrons, in agreement with our observations. Localized MIGSs\cite{Choi2009} are universally present in the interfaces between conductors and insulators, so that this model can be applied to both metallic and semiconducting devices. It is rather intriguing to think that localized MIGSs might well play a key role in both 1/f charge noise, where they provide a trap for electron tunneling to and from the Fermi gas in, say, a SET, and 1/f magnetic flux noise, where they provide localized sites for electrons undergoing spin reversals that couple flux into a SQUID or flux-sensitive qubit.

\hla{Finally, we emphasize that all data acquired at low refrigerator temperature pertain to a voltage-biased, normal-metal SET, implying that the TLFs producing the 1/f charge noise are necessarily out of thermal equilibrium. This is also the case in metrological charge pumps, but not for nondissipative devices, such as charge qubits and Quantum Capacitance Electrometers\cite{Persson2010} (QCEs). It is, however, likely that the microscopic nature of the TLFs is the same in all cases, and a detailed description of the noise in one type of device will surely help the understanding of the noise observed in other devices. Our work implies that one should focus on the nature of the metal-insulator interface to shed light on the nature of the potential wells responsible for the charge noise.}
\begin{acknowledgments}

We are grateful to Vladimir Antonov, Oleg Astafiev, Jonas Bylander, Pierre Echternach, Frank Hekking, John Martinis, Eva Olsson, Kyle Sundqvist, and Dale Van Harlingen for useful discussions, and to Thilo Bauch and Joachim Lublin for assistance with equipment. JC gratefully acknowledges his appointment as Chalmers 150th Anniversary Visiting Professor. The work was supported by the Swedish Research council, the EU project SCOPE and the Wallenberg foundation. This research is based upon work supported in part by the Office of the Director of National Intelligence (ODNI), Intelligence Advanced Research Projects Activity (IARPA) (JC). The views and conclusions contained herein are those of the authors and should not be interpreted as necessarily representing the official policies or endorsements, either expressed or implied, of ODNI, IARPA, or the U.S. Government. The U.S. Government is authorized to reproduce and distribute reprints for Governmental purposes not withstanding any copyright annotation thereon.\end{acknowledgments}

\end{document}